# Relativistic velocity transformation as a genitor of transformation equations (electrodynamics)

Bernhard Rothenstein
Department of Physics, "Politehnica" University of Timisoara, Romania

Aldo De Sabata
Faculty of Electronics and Telecommunications, "Politehnica" University of Timisoara, Romania

*Abstract. We show that invariance of the electric charge and relativistic kinematics lead to the transformation equations for electric field intensity and the magnetic induction.*

## 1. Introduction

Teaching relativistic electrodynamics, the instructor starts with the statement that the electric charge of a particle is a relativistic invariant, i.e. it has the same magnitude for all inertial observers relative to whom it moves with different velocities. The neutrality of an atom inside which the positively charged nucleus and the negatively charged electrons move with very different relative velocities supports the statement.

The approach requires the students to already posses understanding of the elementary notions in relativity (kinematics and dynamics) and classical electrodynamics, specifically:
- An electric field of intensity **E** acts on a particle carrying an electric charge $q$ with an electric force

$$\mathbf{F}_e = q\mathbf{E} \tag{1}$$

which is velocity independent;

-An electric charge $q$ moving with velocity **u** in a magnetic field of magnetic induction **B** detects the action of a magnetic force

$$\mathbf{F}_m = q\mathbf{u} \times \mathbf{B}, \tag{2}$$

**u** representing the velocity of the charged particle relative to the rest frame of the observers who measure **B**.

The purpose of our paper is to derive transformation equations for the physical quantities introduced in order to characterize effects generated by stationary (electrostatics) or moving (electrodynamics) charged particles, mainly based on the composition law of relativistic velocities which can be derived using[1] or avoiding[2] the Lorentz-Einstein transformations for the space-time coordinates of the same event.

## 2. Transformation equations for the components of E and B

The transformation equations for the field quantities **E** and **B** are a starting point for relativistic electrodynamics. Rosser[3] presents two derivations of them. The first one starts with the covariance of Maxwell's equations, involves the Lorentz-Einstein transformations for the space-time coordinates of the same event, which lead to the transformation equations for the partial derivatives with respect to the space and the time coordinates. The second one takes into account the transformation equations for the components of the force and the transformation equations for relativistic velocities as well.

The involved inertial reference frames are $K(XOY)$ and $K'(X'O'Y')$. Their corresponding axes are parallel to each other and the $OX(O'X')$ axes are overlapped. $K'$ moves with velocity **V** relative to $K$ in the positive direction of the overlapped axes. At the origin of time in the two frames ($t=t'=0$) the origins $O$ and $O'$ of the two frames are shortly located at the same point in space. We also consider a third inertial reference frame $K^0(X^0,O^0,Y^0)$ that moves with velocity $u_x$ relative to $K$ but with velocity $u_x'$ relative to $K'$.

The scenario we follow involves a very large conducting surface at rest in $K^0$ reference frame and confined in the $X^0O^0Y^0$ plane. It carries a charge $\sigma^0$ per unit area. The symmetry of the problem indicates that the lines of force are perpendicular to the plane, and if the charge is positive, they are oriented in the positive direction of the $O^0Y^0$ axis. Neglecting side effects, Gauss' theorem tells us that the electric intensity it creates is uniform, has one component:

$$E_y^0 = \frac{\sigma^0}{2\varepsilon_0} = \frac{Q}{2\varepsilon_0 l_{0,x} l_{0,z}}, \qquad (3)$$

and is independent of the distance to the plane. $Q$ represents the charge carried by the plane and $A^0 = l_{0,x} l_{0,y}$ its surface.[4]

Measured by observers from $K$, the same electric field is

$$E_y = \frac{Q}{2\varepsilon_0 l_{0,x}\sqrt{1-\frac{u_x^2}{c^2}} l_{0,z}} \qquad (4)$$

whereas measured by observers from $K'$ it is

$$E_y' = \frac{Q}{2\varepsilon_0 l_{0,x}\sqrt{1-\frac{u_x'^2}{c^2}} l_{0,z}}. \qquad (5)$$

We have taken into account the fact that the true laws of physics are the same in all inertial reference frames, that distances measured perpendicular to the direction of relative motion are relativistic invariants, the Lorentz contraction of lengths measured in the direction of relative motion and the invariance of the vacuum permittivity $\varepsilon_0$. We are interested in a relationship between $E_y$ and $E_y'$, the electric intensities of the same electric field measured by observers from $K$ and $K'$ respectively. We obtain it combining Eqs. (4) and (5), and the result is

$$E_y = E_y' \frac{\sqrt{1-\frac{u_x'^2}{c^2}}}{\sqrt{1-\frac{u_x^2}{c^2}}}. \qquad (6)$$

Eq.(6) becomes a genuine transformation equation if we express its right hand side as a function of physical quantities measured by observers of the $K'$ frame. Because the transformation equation[2] of relativistic velocities relates the velocities $u_x$ and $u_x'$ as

$$u_x = \frac{u_x' + V}{1 + \frac{Vu_x'}{c^2}}, \qquad (7)$$

(6) leads to

$$E_y = E_y' \frac{1 + \frac{Vu_x'}{c^2}}{\sqrt{1-\frac{V^2}{c^2}}} = \frac{E_y' + \frac{V}{c^2} E_y' u_x'}{\sqrt{1-\frac{V^2}{c^2}}} \qquad (8)$$

and thus we have derived the transformation for the $OY(O'Y')$ components of the electric intensity.

We detect in the right hand side of (8) the presence of the term

$$\frac{u_x'}{c^2} E_y' \qquad (9)$$

which reads in $K$

$$\frac{u_x}{c^2} E_y. \qquad (10)$$

Physicists are well-trained godfathers finding out names not only for basic physical quantities they use in order to build a system of units (length, mass, time, absolute temperature and electric charge) but for combinations of them as well (speed, acceleration, momentum, energy, force, pressure and many others). They introduce the notations

$$B_z = \frac{u_x E_y}{c^2} \tag{11}$$

in $K$ and

$$B'_z = \frac{u'_x E'_y}{c^2} \tag{12}$$

in $K'$, considering that it has the character of a vector, showing in the positive direction of the $OZ(O'Z')$ axis in such a way that the cross products

$$\mathbf{u}_x \times \mathbf{E}_y \tag{13}$$

$$\mathbf{u}'_x \times \mathbf{E}'_y \tag{14}$$

show in the same direction as $\mathbf{B}_z$ and $\mathbf{B}'_z$ show. They call $B_z$ and $B'_z$ the $OZ(O'Z')$ components of the magnetic induction. It transforms as

$$B_z = \frac{u_x E_y}{c^2} = B'_z \frac{1 + \dfrac{V}{u'_x}}{\sqrt{1 - \dfrac{V^2}{c^2}}} = \frac{B'_z + \dfrac{V}{c^2} E'_y}{\sqrt{1 - \dfrac{V^2}{c^2}}}. \tag{15}$$

Confining the uniformly charged plane we have considered so far, in the $X^0 O^0 Y^0$ plane it creates a uniform electric field $E_{0,z}$ pointing into the positive direction of the $OZ$ axis. Measured by observers from $K$, it is $E_z$ whereas for those of the $K'$ frame it is $E'_z$. Moving, it generates the magnetic inductions $B_y$ in $K$ and $B'_y$ in $K'$, both showing in the negative direction of the $OY(O'Y')$ axes. The same strategy as above leads to the transformations

$$E_z = \frac{E'_z - V B'_y}{\sqrt{1 - \dfrac{V^2}{c^2}}} \tag{16}$$

$$B_y = \frac{B'_y - \dfrac{V}{c^2} E'_z}{\sqrt{1 - \dfrac{V^2}{c^2}}}. \tag{17}$$

Consider now that the charged plane is confined in the $YOZ(Y'O'Z')$ plane, generating a uniform electric field that shows in the positive direction of the overlapped axes $OX(O'X')$, $E_x$ and $E'_x$ respectively, given by

$$E_x = \frac{\sigma}{2\varepsilon_0} = \frac{Q}{2\varepsilon_0 l_{0,y} l_{0,z}} \tag{18}$$

$$E'_x = \frac{\sigma'}{2\varepsilon_0} = \frac{Q}{2\varepsilon_0 l'_{0,y} l'_{0,z}}. \tag{19}$$

All the physical quantities in the right hand side of (18) and (19) are relativistic invariants, which leads to

$$E_x = E'_x. \tag{20}$$

The motion of the charged plane does not generate a magnetic field.
For deriving the transformation equations for the $OX(O'X')$ components of the magnetic field, we consider a solenoid composed of several coaxial loops, all carrying the same current. Its axis coincides with the overlapped axes $OX(O'X')$. All the loops carry the same current. In its rest frame, the solenoid generates a magnetic induction showing in the positive direction of the overlapped axes given by

$$B_{0,x} = \frac{\mu_0 N}{2l_{0,x}} \frac{dQ}{dt_0}, \quad (21)$$

where $N$ represents the invariant number of loops and $I_0 = \frac{dQ}{dt_0}$ represents the current carried by each loop, $dt_0$ representing a proper time interval. When detected from another reference frame, $l_{0,x}$ contracts and $dt_0$ dilates resulting that the $OX(O'X')$ component of the magnetic induction is a relativistic invariant

$$B_x = B'_x.$$

As compared with other derivations of the transformations for **E** and **B**, we can consider that our derivation is a two lines one. The main relativistic ingredient we used is the transformation of relativistic velocities. The strategy we have followed deriving the transformations for **E** and **B** leads to the transformation equations for charge and current densities.

For those learners who define the magnetic induction **B** via the magnetic force

$$\mathbf{f} = q\mathbf{u} \times \mathbf{B} \quad (22)$$

as

$$B = \frac{(f_{mag})_{max}}{qu} \quad (23)$$

we present a derivation for the components of **E** and **B** based on the transformation of relativistic velocities.

Consider a particle that carries a positive charge $q$. It enters with velocity $u_x$ in a space where the observers of the $K$ frame detect an electric field intensity $E_y$ and a magnetic induction $B_z$. If the condition

$$qE_y = qu_x B_z \quad (24)$$

is fulfilled, the particle continues to move along the $OX$ axis. The principle of relativity tells us that (24) reads in $K'$

$$qE'_y = qu'_x B'_z \quad (25)$$

and the particle moves in that frame along the $O'X'$ axis as well.

Combining (24) and (25) we obtain

$$\frac{E_y}{B_z} = \frac{E'_y}{B'_z} \frac{u_x}{u'_x} = \frac{E'_y}{B'_z} \frac{1+\frac{V}{u'_x}}{1+\frac{Vu'_x}{c^2}}. \quad (26)$$

Eq.(26) suggests to consider that

$$E_y = f(V) E'_y (1+\frac{V}{u'_x}) \quad (27)$$

$$B_z = f(V) B'_z (1+\frac{Vu'_x}{c^2}) \quad (28)$$

where $f(V)$ represents an unknown function of the relative velocity $V$, but not of the physical quantities involved in the transformation process (linearity of the transformation equations).
Combing (23) and (24) yields

$$\frac{E'_y}{B'_z} = \frac{E_y}{B_z} \frac{u'_x}{u_x} = \frac{E_y}{B_z} \frac{1-\frac{V}{u_x}}{1-\frac{Vu_x}{c^2}}. \quad (29)$$

Eq.(29) suggests to consider that

$$E'_y = f(V)E_y(1-\frac{V}{u_x}) \qquad (30)$$

$$B'_z = f(V)B_z(1-\frac{Vu_x}{c^2}) \qquad (31)$$

with the same function $f(V)$, taking into account the principle of reciprocity. By multiplying (28) and (31) side by side we obtain

$$f^2(V)(1+\frac{Vu'_x}{c^2})(1-\frac{Vu_x}{c^2}) = 1. \qquad (32)$$

Expressing the left hand side of (32) as a function of $u_x$ or as a function of $u'_x$ via the transformation equation of relativistic velocities we obtain for $f(V)$

$$f(V) = \gamma(V) = \frac{1}{\sqrt{1-\frac{V^2}{c^2}}} \qquad (33)$$

and the transformation equations we are looking for are

$$E_y = \gamma(V)E'_y(1+\frac{V}{u'_x}) = D_{E_y}(V,u'_x)E'_y \qquad (34)$$

$$B_z = \gamma(V)B'_z(1+\frac{Vu'_x}{c^2}) = D_{B_z}(V,u'_x)B'_z \qquad (35)$$

Consider[5] a plane acoustic wave propagating with velocity $u_x$ relative to $K$ and with velocity $u'_x$ relative to $K'$ in the positive direction of the overlapped axes. Imposing the condition of invariance to the phase, the result is that the frequency of the oscillations taking place in the wave ($\upsilon$ in $K$ and $\upsilon'$ in $K'$) transform as

$$\nu = \gamma(V)\nu'(1+\frac{V}{u'_x}) = D_\nu(V,u'_x)\nu'. \qquad (36)$$

The wave number ($k$ in $K$ and $k'$ in $K'$) transforms as

$$k_x = \gamma(V)k'_x(1+\frac{Vu'_x}{c^2}) = D_{k_x}(V,u'_x)k'_x. \qquad (37)$$

Comparing (34) and (36) we see that the $OY(O'Y')$ components of **E** transform as the frequency in an acoustic wave does. Comparing (35) and (37) we see that the $OZ(O'Z')$ components of **B** transform as the wave vector in an acoustic wave does.
The transformation factors $D_{E_y}(V,u'_x)$, $D_{B_z}(V,u'_x)$, $D_\nu(V,u'_x)$, and $D_k(V,u'_x)$ share the property that, for $u_x = u'_x = c$, they become equal to each other, i.e.

$$D_{E_y}(V,c) = D_\nu(V,c) = D_{B_z}(V,c) = D_{k_x}(V,c) \qquad (38)$$

**3. The plane electromagnetic wave**
Consider a plane electromagnetic wave propagating in the positive direction of the overlapped axes. If $\nu_c$, $\nu'_c$, $k_{x,c}$, and $k'_{x,c}$ represent the frequencies of the electromagnetic oscillations taking place in the wave and the wave vectors as detected from $K$ and $K'$ respectively, then they transform as

$$k_{x,c} = D_{k_x,c}(V)k'_{x,c} \qquad (39)$$

$$\nu_c = D_{\nu,c}(V)\nu'_c. \qquad (40)$$

If special relativity ensures a smooth transition from the case of subluminal motion to the case of luminal propagation, then we should have for the electric and the magnetic components of the electromagnetic oscillations

$$E_{y,c} = \gamma(V)(1+\frac{V}{c})E'_{y,c} \tag{41}$$

$$B_{z,c} = \gamma(V)(1+\frac{V}{c})B'_{z,c} . \tag{42}$$

Eq.(11) tells us that the magnitudes of **E**$_c$ and **B**$_c$ are related by

$$B_{z,c} = \frac{E_{y,c}}{c} . \tag{43}$$

**4. Conclusions**

The traditional way to derive the transformation equations for the components of **E** and **B** use the Lorentz-Einstein transformations for the space-time coordinates and the transformation of force as well. Our derivation uses only the relativistic velocity transformation and the invariance of the electric charge. The usual transformation equations present the $E_y$ component as a function of $E'_y$ and of $B'_z$ as well. Our derivation presents $E_y$ as a function of $E'_y$ and of a transformation factor that depends on $V$ and on $u'_x$ (15). We find the same situation in the case of the transformation of the component of **B**.

Our derivation shows clearly that a magnetic field is no more then a moving electric field and does not make use of Maxwell's equations.